\documentclass{appolb}
\usepackage{graphicx, color, subfigure, slashed, listings, fancyhdr, amsmath, amsthm, amssymb, bbm}

\begin{document}
\title{Towards a self-consistent solution of the Landau gauge quark-gluon vertex Dyson-Schwinger equation %
\thanks{Presented at Light Cone 2012, $8^{th}$ -- $13^{th}$ July 2012, Polish Academy of Sciences, Cracow, Poland}%
}
\author{Andreas Windisch, Markus Hopfer and Reinhard Alkofer
\address{Institut f\"ur Physik, Karl-Franzens Universit\"at Graz\\ Universit\"atsplatz 5, 8010 Graz, Austria}
}
\maketitle
\begin{abstract}
The quark-gluon vertex in Landau gauge QCD is investigated 
in the Dyson-Schwinger approach. The aim is to obtain a fully self-consistent solution 
of the quark propagator and the quark-gluon vertex Dyson-Schwinger equation (DSE) in the 
vacuum using the gluon propagator as input from other calculations.
The truncation scheme used in an earlier study is systematically improved and the first decisive step 
towards a full solution is presented. First results for the quark propagator with a fully 
dressed quark-gluon vertex in a truncation that incorporates all twelve vertex 
tensor-structures are shown.
\end{abstract}
\PACS{11.15.-q, 11.30.Rd, 12.38.Aw}
\section{Motivation}
Prominent features of QCD like confinement and dynamical chiral symmetry breaking (D$\chi$SB)  
are still not satisfactorily  
understood. One possibility to explore the non-per\-tur\-ba\-tive regime of QCD is to employ 
DSEs, an infinite tower of coupled 
integral equations, see {\it e.g.}, ref.\ \cite{Alkofer:2000wg} and references therein. 
Their non-perturbative nature makes them an appropriate tool for exploring 
the low-energy domain of the theory. Careful truncations have to be imposed on the system in order
to evaluate the equations. In these proceedings we will extend a prior study of the
quark-gluon vertex \cite{Alkofer:2008tt} by systematically 
improving on the  used truncation. The aim is to shed more light on a
possible relation between confinement and D$\chi$SB by studying the quark-gluon
vertex.\par
The starting point is the system of equations for the quark propagator and the quark-gluon vertex
in Landau gauge using fits for the gluon propagator as input \cite{Alkofer:2008tt,Fischer:2002hna}. 
The three-gluon vertex is modeled  as in ref.~\cite{Alkofer:2008tt}.\footnote{These fits relate to the so-called 
scaling solution, one of two types 
of solutions for Landau gauge Green functions when classified by their infrared behavior, see {\it e.g.}, refs.\ 
\cite{Fischer:2008uz,LlanesEstrada:2012my} and references therein. However, we want to emphasize here 
the evidence that the difference between these types of solutions is irrelevant for phenomenology.}
The infrared suppression of the gluon propagator relates on the one hand to an infrared enhancement of the ghost 
propagator (see {\it e.g.}, ref.\ \cite{Watson:2001yv}) and on the other hand to the positivity violation for 
transverse gluons. The latter property indicates the role of the transverse gluons in a non-perturbative 
BRST quartet mechanism \cite{Alkofer:2011pe} and therefore a certain aspect of gluon confinement: In Landau 
gauge QCD transverse gluons are confined and therefore not confining.
Thus, given the infrared suppression of the gluon propagator,  the quark-gluon vertex has to
acquire a certain strength in the infrared to provide D$\chi$SB and, eventually, quark confinement.
In ref.\ \cite{Alkofer:2008tt} it has been found that within the imposed truncation (and at vanishing temperatures 
and densities)  D$\chi$SB
and quark confinement occur either together in a self-consistent way or are both absent.
 By improving on the truncation of the quark-gluon vertex
DSE we  hope to get deeper insights into the possible relation between quark confinement and
D$\chi$SB. In addition, isolating important tensor structures of the quark-gluon vertex 
will likely lead to improved phenomenological applications. Especially, corresponding studies 
at non-vanishing temperatures and/or quark chemical potentials
can benefit from a
more detailed understanding of the quark-gluon vertex, see {\it e.g.}, ref.~\cite{Hopfer:2012lc}.    
\section{The coupled system}
\begin{figure}[h]
\centerline{%
\includegraphics[width=12.5cm]{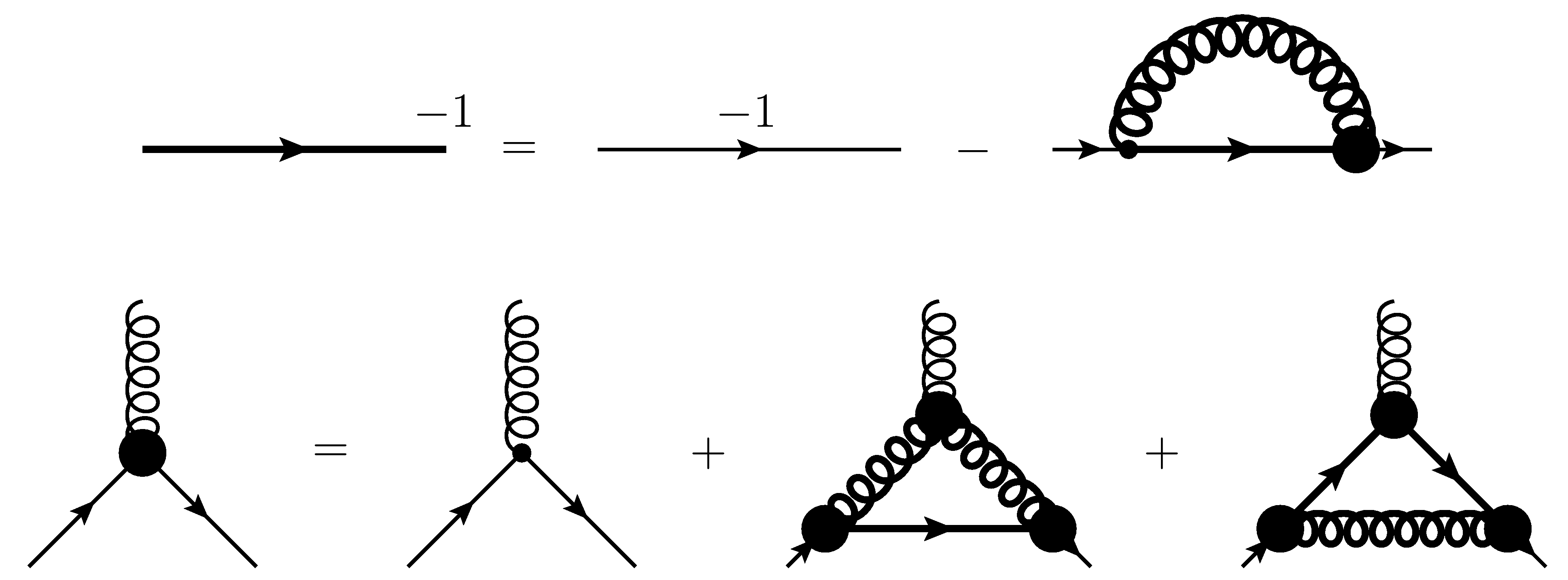}}
\caption{The quark propagator and the quark-gluon vertex DSE.
All internal propagators are dressed. The vertex equation has been derived from a 
3PI effective action \cite{Berges:2004pu}.}
\label{Fig1}
\end{figure}

The coupled system of the quark propagator and the quark-gluon vertex is shown in 
Fig. \ref{Fig1}. This system serves as a starting point for this study.
The vertex equation in Fig. \ref{Fig1} has been derived from a 3PI effective action
and corresponds to the DSE truncated at three-point level \cite{Berges:2004pu}. Note that 
as usual in an nPI-based approach all vertices are dressed.
For obvious reasons, the last two diagrams on the right-hand side of the vertex equation are 
referred to as the \textit{non-Abelian} and the \textit{Abelian} diagram. Upon performing the 
color traces one finds that the non-Abelian diagram has a prefactor of ${N_c}/{2}$, whereas
the Abelian diagram is suppressed by a factor of $-{1}/{2N_c}$. This suppression has been found
to be even more severe due to the dynamics of the system \cite{Alkofer:2008tt}. Thus, as a first step
to solve the system of the quark propagator and the quark-gluon vertex DSE, only the non-Abelian diagram
has been taken into account. 
\subsection{A first step}
\label{sec:a_first_step}
\begin{figure}[h]
\centerline{%
\includegraphics[width=10cm]{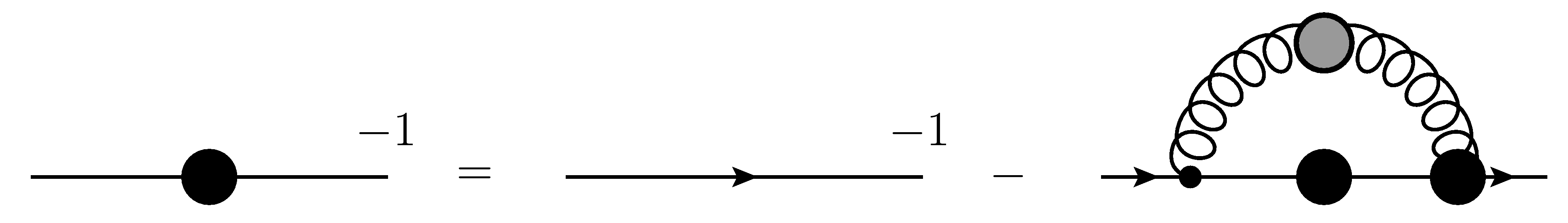}}
\caption{Quark propagator DSE. Full blobs denote fully dressed quantities, shaded blobs are fits 
from other studies.}
\label{Fig2}
\end{figure}

\begin{figure}[h]
\centerline{%
\includegraphics[width=10cm]{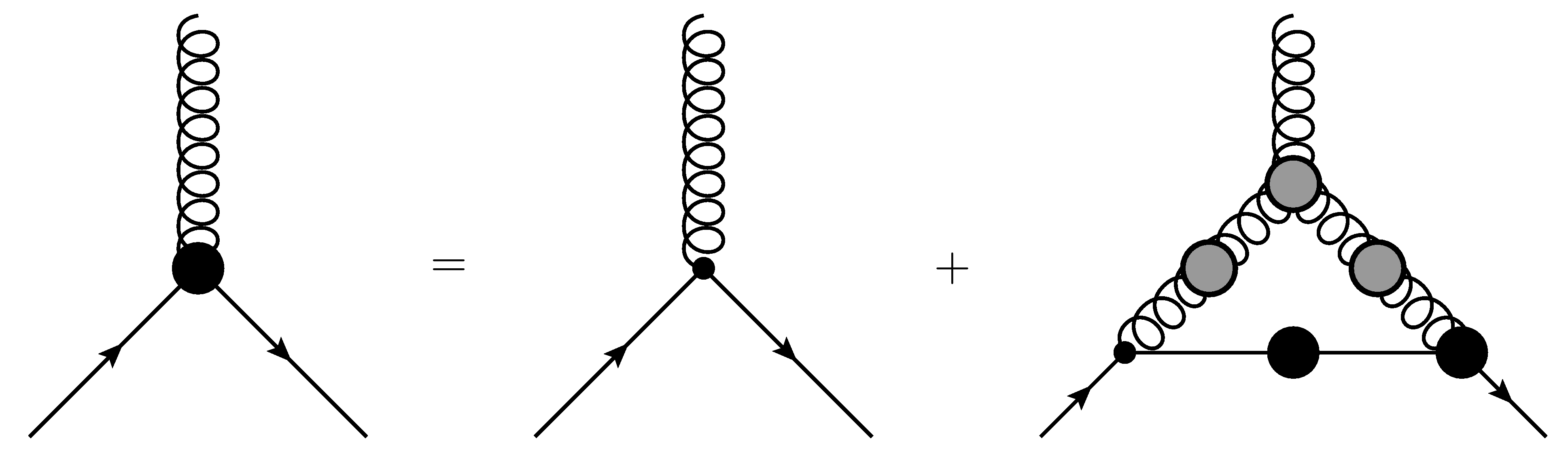}}
\caption{The truncated quark-gluon vertex DSE. Full blobs denote fully 
dressed quantities, shaded blobs are fits from other studies. 
For the three-gluon vertex a model has been used which is adopted from \cite{Alkofer:2008tt}.}
\label{Fig3}
\end{figure}

The system as depicted in Figs.\ \ref{Fig2} and  \ref{Fig3} has been considered as a first step towards
the full solution of the system of Fig. \ref{Fig1}.   Within this approximation,
 one vertex of the non-Abelian diagram remains bare while the second
is dressed such that the full vertex experiences feed-back from all twelve 
tensor structures. This vertex also enters the propagator equation Fig. \ref{Fig2}. In the vacuum, 
twelve tensor structures are needed as a basis to span the vertex. Instead of employing 
the widely-used Ball-Chiu basis \cite{Ball:1980ay} it turned out to be numerically advantageous 
to use a simpler basis given by
\begin{equation}
\Gamma^\mu(p_1,p_2)\propto\left\{\begin{array}{c}
\mathbbm{1}\\
\slashed p_1\\
\slashed p_2\\
\frac{1}{2}{[\slashed p_1,\slashed p_2 ]}
\end{array}\right\}\otimes\left\{\begin{array}{c}
\gamma^\mu\\
p_1^\mu\\
p_2^\mu
\end{array}\right\}\nonumber
\end{equation}
where $p_1$ and $p_2$ are the in- and outgoing quark momenta. 
As input from the
Yang-Mills sector the gluon propagator has been taken from earlier DSE studies\footnote{Here, 
a scaling-type fit obtained from a self-consistent solution of the corresponding ghost-gluon system
has been used \cite{Fischer:2002hna}. The fit parameter values are taken from \cite{Alkofer:2008tt}.}. 
The three-gluon vertex is represented by its tree-level tensor structure combined with the following
ansatz \cite{Alkofer:2008tt}
\begin{equation}
\Gamma^{3g}(x)=\left(\frac{x}{d_1+x}\right)^{-3\kappa}
\left(d_3\frac{d_1}{d_1+x}+d_2\log\biggl[\frac{x}{d_1}+1\biggr]\right)^{17/44}
\end{equation}
where $x=p_1^2+p_2^2+p_3^2$ is the sum over the squared gluon momenta entering the vertex and 
$\kappa \approx 0.595$. 
This ansatz reflects the strong IR enhancement known 
from Yang-Mills theory (see, however, the remark in sect.~3) 
as well as ensures the correct running in the UV region. The residual parameters
$d_1$, $d_2$ and $d_3$ have to be fixed to physical observables, e.g. the chiral condensate.
\subsection{Obtaining the vertex dressing functions}
As a first step, the dressing functions have to be disentangled to get them
in an explicit form. Since the employed basis is neither orthonormal 
nor orthogonal one obtains a linear system of equations for the vertex dressing functions in terms of the 
twelve right-hand side projections. This system can be solved in advance, giving rise to explicit
expressions for the dressing functions as linear combinations of the right-hand side
 projections\footnote{The algebraic manipulations have been carried out with the program 
\texttt{FORM} \cite{Vermaseren:2000nd}.}.
In each iteration step these projections are evaluated
numerically and are used subsequently to obtain the dressing functions by inserting them into the solution
of the pre-calculated linear system of equations.  
\subsection{Renormalization and numerical treatment}
The quark wave function renormalization constant $Z_2$, the mass renormalization constant $Z_m$
 as well as the quark-gluon vertex
renormalization constant $Z_{1F}$ has been fixed within a MOM scheme. 
In particular, the renormalization conditions for the quark dressing function $A(\mu^2)=1$ and 
the tree-level vertex dressing function $\lambda_1(\mu^2,\mu^2,2\mu^2)=1$
have been employed, where $\mu^2=170\,$GeV$^2$ denotes the renormalization scale.
Only the tree-level tensor $\gamma^\mu$ has to be concerned in this procedure since
all other tensor structures lead to UV finite expressions. 
$Z_m$ has been fixed by the condition $B(\mu^2)=m_R$, 
where $m_R$ is the mass at the renormalization scale.

The evaluation of the coupled system of the quark propagator and the quark-gluon
vertex as depicted in Fig. \ref{Fig2} and Fig. \ref{Fig3} is numerically demanding. 
The actual calculations are thus performed on a cluster featuring Graphics
Processing Units (GPUs). GPUs have already been applied successfully to solve DSEs
\cite{Hopfer:2012ht}. They will be vital once the full non-Abelian diagram
and/or the Abelian diagram is taken into account.
\section{First results}
\begin{figure}[h]
\centering
\subfigure{\includegraphics[width=6.28cm]{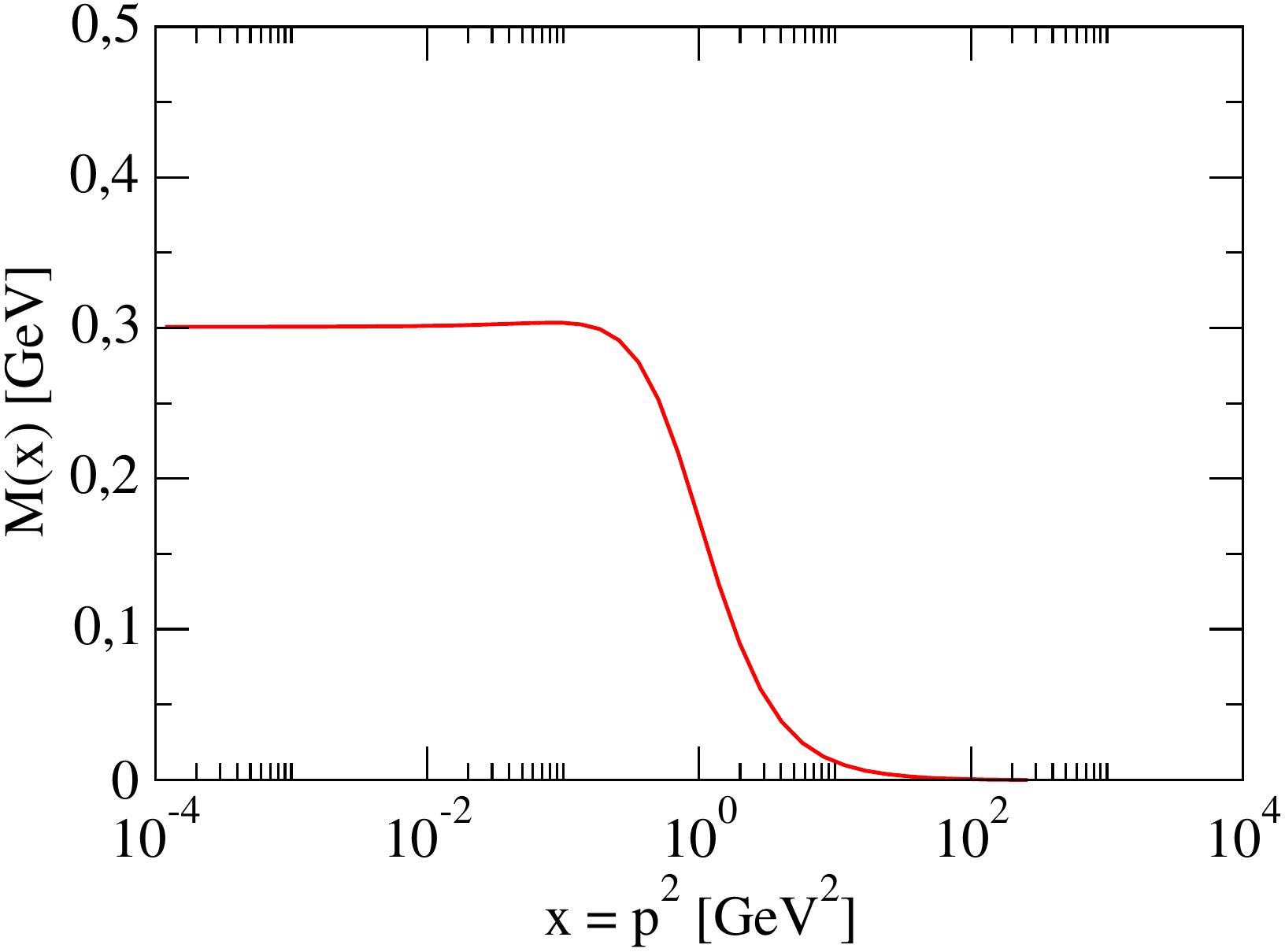}\label{Fig4a}}
\subfigure{\includegraphics[width=6.2cm]{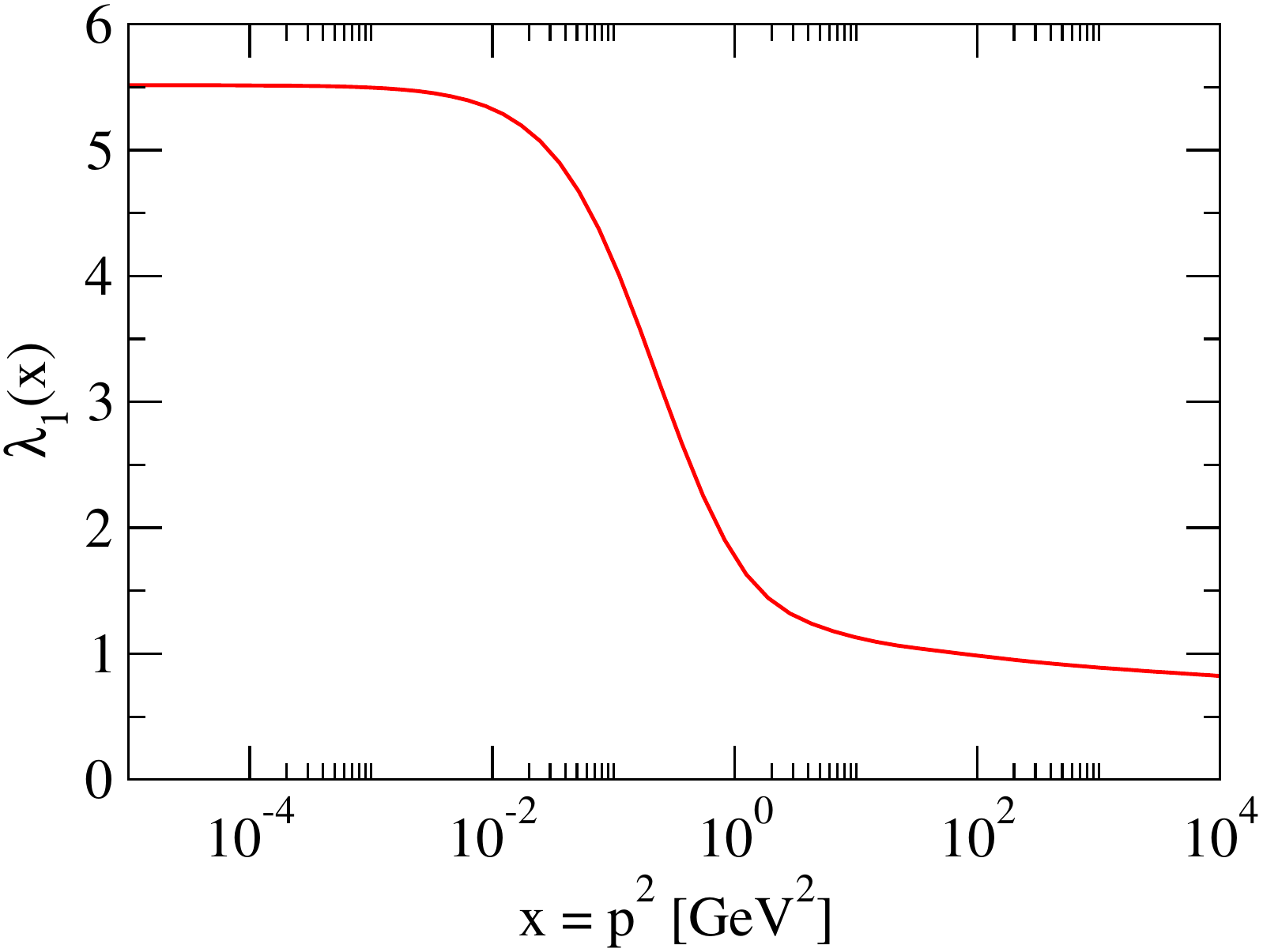}\label{Fig4b}}
\caption{Left: The mass function $M(p^2)=B(p^2)/A(p^2)$ as obtained from the coupled system
described in Sec. \ref{sec:a_first_step}. Right: The vertex dressing function $\lambda_1(p^2)$ 
of the tree-level part $\gamma^\mu$ of the quark-gluon vertex evaluated at $p_1^2=p_2^2=p^2$ and $p_1.p_2=0$.}
\end{figure}

The system described in Sec. \ref{sec:a_first_step} has been solved self-consistently, taking all twelve
tensor structures for the quark-gluon vertex into account. Fig. \ref{Fig4a} shows the behaviour of the
mass function $M(p^2)$ as obtained from the quark propagator dressing functions in the chiral limit.
Due to stability issues it was not yet possible to take the
full IR strength of the three-gluon vertex $(p^2)^{-3\kappa}$ into account\footnote{In our calculations the
corresponding IR exponent has been varied from $0$ to $-2\kappa$ resulting in moderate changes of the 
dynamical generated mass.}.
Fig. \ref{Fig4b} shows the vertex dressing function $\lambda_1$ which corresponds to the
tree-level structure $\gamma^\mu$ evaluated at the symmetric point $p_1^2=p_2^2=p^2$ and $p_1.p_2=0$.
The significant IR enhancement seen in this function is also present in most other tensor structures, independent 
whether they have been generated by D$\chi$SB or are chirally symmetric.

\section{Conclusions and outlook}
The coupled DSEs for the quark propagator and the 
quark-gluon vertex in Landau gauge has been investigated. 
Results for the quark propagator coupled to a partially dressed quark-gluon vertex including all
twelve tensor structures have been presented.
Isolating the 
important tensor structures will become vital for subsequent calculations taking the full non-Abelian
as well as the Abelian diagram into account, aiming towards a complete solution of the system as depicted in
Fig.~\ref{Fig1}. Such a reduction of complexity is also mandatory in studies involving 
the quark-gluon vertex, especially those at non-vanishing temperatures and/or quark chemical potentials.
\section*{Acknowledgments}
We thank Gernot Eichmann, Christian Fischer, Markus Q. Huber, Manfred Liebmann, Felipe Llanes-Estrada, 
Mario Mitter and Richard Williams for helpful discussions. MH and AW acknowledge support 
by the Doktoratskolleg ''Hadrons in Vacuum, Nuclei and Stars`` of the Austrian Science Fund, 
FWF DK W1203-N16. This study is also supported by the 
Research Core Area ''Modeling and Simulation`` of the University of Graz, Austria.  

\end{document}